# Fermi liquid behavior of the in-plane resistivity in the pseudogap state of YBa$_2$Cu$_4$O$_8$

Cyril Proust[1*], B. Vignolle[1], J. Levallois[1,2], S. Adachi[3] and N. E. Hussey[4,5*]

[1] *Laboratoire National des Champs Magnétiques Intenses (LNCMI-EMFL), (CNRS-INSA-UGA-UPS), Toulouse 31400, France.*

[2] *Department of Quantum Matter Physics, University of Geneva, CH-1211 Geneva 4, Switzerland.*

[3] *Superconductivity Research Laboratory, Shinonome 1-10-13, Koto-ku, Tokyo 135-0062, Japan.*

[4] *High Field Magnet Laboratory (HFML-EMFL), Radboud University, Toernooiveld 7, 6525ED Nijmegen, Netherlands*

[5] *Radboud University, Institute of Molecules and Materials, Heyendaalseweg 135, 6525 AJ Nijmegen, Netherlands*

[*] To whom correspondence may be addressed. Email: cyril.proust@lncmi.cnrs.fr or N.E.Hussey@science.ru.nl

.

Our knowledge of the ground state of underdoped hole-doped cuprates has evolved considerably over the last few years. There is now compelling evidence that inside the pseudogap phase, charge order breaks translational symmetry leading to a reconstructed Fermi surface made of small pockets. Quantum oscillations, (Doiron-Leyraud N, et al. (2007) Nature 447:564-568), optical conductivity (Mirzaei SI, et al. (2013) Proc Natl Acad Sci USA 110:5774-5778) and the validity of Wiedemann-Franz law (Grissonnache G, et al. (2016) Phys. Rev. B 93:064513) point to a Fermi liquid regime at low temperature in the underdoped regime. However, the observation of a quadratic temperature dependence in the electrical resistivity at low temperatures, the hallmark of a Fermi liquid regime, is still missing. Here, we report magnetoresistance measurements in the magnetic-field-induced normal state of underdoped $YBa_2Cu_4O_8$ which are consistent with a $T^2$ resistivity extending down to 1.5 K. The magnitude of the $T^2$ coefficient, however, is much smaller than expected for a single pocket of the mass and size observed in quantum oscillations, implying that the reconstructed Fermi surface must consist of at least one additional pocket.

## Significance

High temperature superconductivity evolves out of a metallic state that undergoes profound changes as a function of carrier concentration, changes that are often obscured by the high upper critical fields. In the more disordered cuprate families, field suppression of superconductivity has uncovered an underlying ground state that exhibits unusual localization behavior. Here, we reveal that in stoichiometric $YBa_2Cu_4O_8$, the field-induced ground state is both metallic and Fermi-liquid like. The manuscript also demonstrates the potential for using the absolute magnitude of the electrical resistivity to constrain the Fermi surface topology of correlated metals and in the case of $YBa_2Cu_4O_8$, reveals that the current picture of the reconstructed Fermi surface in underdoped cuprates as a single, isotropic electron-like pocket may be incomplete.

## Introduction

The generic phase diagram of Fig. 1 summarizes the temperature and doping dependence of the in-plane resistivity $\rho_{ab}(T)$ of hole-doped cuprates[1]. Starting from the heavily overdoped side, non-superconducting $La_{1.67}Sr_{0.33}CuO_4$ for example shows a purely quadratic resistivity below $\sim 50$ K (ref. 2). Below a critical doping $p_{SC}$ where superconductivity sets in, $\rho_{ab}(T)$ exhibits supra-linear behaviour that can be modeled either as $\rho_{ab} \sim T + T^2$ or as $T^n$ ($1 < n < 2$). When a magnetic field is applied to suppress superconductivity on the overdoped side, the limiting low-$T$ behavior is found to be $T$-linear[3,4,5]. Optimally doped cuprates are characterized by a linear resistivity for all $T > T_c$, though the slope often extrapolates to a negative intercept suggesting that at the lowest temperatures, $\rho_{ab}(T)$ contains a component with an exponent larger than one[1]. In the underdoped regime, $\rho_{ab}(T)$ varies approximately linearly with temperature at high $T$, but as the temperature is lowered below the pseudogap temperature $T^*$, it deviates from linearity in a very gradual way[6]. At lower temperatures, marked by the light blue area in Fig. 1, there is now compelling evidence from various experimental probes of incipient charge order[7-11]. High field NMR[12,13] and ultrasonic[14] measurements indicate that a phase transition occurs below $T_c$. This is also confirmed by recent high field x-ray measurements which indicate that the CDW order becomes tridimensional with a coherence length that increases with increasing magnetic field strength[15,16]. This leads to a Fermi surface reconstruction that can be reconciled with quantum oscillations[17,18] as well as with the sign change of the Hall[19] and Seebeck[20] coefficients. Whether the charge order is biaxial[21] or uniaxial with orthogonal domains[22] is still an open issue, but a Fermi surface reconstruction involving two perpendicular wavevectors leads to at least one electron pocket in the nodal region of the Brillouin zone[23]. Depending on the initial pseudogapped Fermi surface and on the wavevectors of the charge order, Fermi surface reconstruction can also lead to additional, smaller hole pockets[24,25]. In Y123 (doping level p ≈ 0.11) the quantum oscillation spectra consist of a main frequency $F_a = 540$ T and a beat pattern indicative of nearby frequencies $F_{a2} = 450$ T and $F_{a3} = 630$ T. A smaller frequency $F_h ≈ 100$ T has been detected by thermopower and c-axis transport measurements and attributed there to an additional small hole pocket[26]. The presence of the three nearby frequencies $F_{ai}$ can be explained by a model involving a bilayer system with an electron pocket in each plane and magnetic breakdown between the two pockets[27,28]. In this scenario, the low frequency $F_h$ could originate from quantum interference or the Stark effect[29]. However, this scenario predicts the occurrence of five nearby frequencies, and thus requires fine tuning of certain microscopic parameters such as the bilayer tunneling $t_\perp$. Moreover, the doping dependence of the Seebeck coefficient is difficult to reconcile with a Fermi surface reconstruction scenario leading to only one electron pocket per plane.

More generally, the observation of quantum oscillations is a classic signature of Landau quasiparticles. In underdoped Y123, the temperature dependence of the amplitude of the oscillations follows Fermi-Dirac statistics up to 18 K, as in the Landau-Fermi liquid theory[30]. This conclusion is supported by other observations, such as the validity of the Wiedemann-Franz law[31] in underdoped Y123 and the quadratic frequency and temperature dependence of the quasiparticle lifetime $\tau(\omega, T)$ measured by optical spectroscopy[32] in underdoped HgBa$_2$CuO$_{4+\delta}$ (Hg1201). An important outstanding question is whether the in-plane resistivity of underdoped cuprates also exhibits the behavior of a canonical Landau-Fermi liquid, namely a quadratic temperature dependence at low $T$? In underdoped cuprates, several studies have shown $\rho_{ab} \sim T^2$, but always at elevated temperatures (indeed, most are above $T_c$) and only over a limited temperature range that never exceed a factor of 2.5[6,33 -36]. (see Table S1 for a detailed list of studies). At low temperatures, either $\rho_{ab}(T)$ starts to become non metallic[36], suggesting that the $T^2$ behavior observed at intermediate temperatures could just be a crossover regime, or a quadratic behaviour has previously been hinted at[19], rather than shown explicitly. Here, we present high field in-plane magnetoresistance measurements in underdoped YBa$_2$Cu$_4$O$_8$ (Y124) that are consistent with the form $\rho_a(T) = \rho_0(T) + AT^2$ from $T = T_c$ down to temperatures as low as 1.5 K, e.g. over almost two decades in temperature. In addition, we investigate the magnitude of the resultant $A$ coefficient and compare it with some of the prevalent Fermi surface reconstruction scenarios. In conclusion, we find that the magnitude of $A$ is difficult to reconcile with the existence of a single electron pocket per plane, with an isotropic mass.

## Results

We have measured the a-axis magnetoresistance (i.e. perpendicular to the conducting CuO chains) of two underdoped Y124 ($T_c = 80$ K) single crystals up to 60 T at various fixed temperatures down to 1.5 K. From thermal conductivity measurements at high fields, the upper critical field $H_{c2}$ of Y124 has been estimated to be ~ 45 T[37]. Raw data for both samples are shown in Fig. 2. Above $T_c$, i.e. in the absence of superconductivity, the transverse magnetoresistance can be accounted for, *over the entire field range measured*, by a two-carrier model[35] using the formula:

$$\rho(H) = \rho(0) + \frac{\alpha H^2}{1 + \beta H^2} \qquad (1)$$

where $\rho(0)$ is the zero-field resistivity and $\alpha$ and $\beta$ are free parameters that depend on the conductivity and the Hall coefficient of the electron and hole carriers[35] (see Fig. S1 for a comparison of the two-band and single-band, quadratic forms for the magnetoresistance.) To obtain reliable values of $\rho(H\rightarrow 0, T) = \rho(0)$ and corresponding error bars for each field sweep, the data were fitted to equation (1) in varying field ranges using the procedure described in detail in Figs. S2-S5 and Tables S2-S3. Precisely the same form is used to fit the high-field data at all temperatures studied below $T_c$. This procedure has been found to yield reliable $\rho(0)$ values in both cuprate[5] and pnictide[38] superconductors. Extrapolation of the high-field data to the zero-field axis $\rho(0)$, as shown by dashed lines in Fig. 2, allows one then to follow the evolution of $\rho_a(T)$ down to low temperatures. The extrapolated $\rho(0)$ values are plotted versus temperature in Fig. 3 (symbols) for both crystals, along with the zero-field temperature dependence of the resistivity (solid line). The dashed lines in Fig. 3 correspond to fits of the $\rho(0, T)$ data to the form $\rho_0(T) = \rho_0 + AT^2$. A fit of the data to the form $\rho_0(T) = \rho_0 + AT^n$ yields $n = 1.9 \pm 0.2$ is shown in the Fig. S6. The inset of the Fig. 3 are corresponding plots of $\rho_a(T)$ versus $T^2$ to highlight the approximately quadratic form of $\rho_a(T)$. From the dashed line fits we obtain $\rho_0 = 7.5 \pm 1.0$ and $10.0 \pm 1.0$ $\mu\Omega$.cm and $A = 10.0 \pm 1$ and $8.5 \pm 0.5$ n$\Omega$ cm K$^{-2}$ for samples #1 and #2 respectively.

## Discussion

The first key result of this study is our observation, within experimental resolution, of a quadratic temperature dependence of the in-plane resistivity in Y124 down to low temperatures, which indicates that the low-lying (near-nodal) electronic states inside the pseudogap phase of underdoped cuprates bear all the hallmarks of Landau quasiparticles. Intriguingly, the magnitude of the $T^2$ term, $A = 9.5 \pm 1.5$ n$\Omega$ cm K$^{-2}$, is similar to that measured at high temperature, e.g. above 50 K in underdoped Y123 at $p = 0.11$ ($A \approx 6.6$ n$\Omega$ cm K$^{-2}$) (ref. 33) as well as in single-layer Hg1201 above 80 K where $A$ varies between $\approx 10$ and 15 n$\Omega$ cm K$^{-2}$ for $0.055 \le p \le 0.11$ (ref. 34). Note that in Y124, such comparison cannot be made since the temperature dependence of the resistivity is not quadratic above $T_c$. In Y123, it is known that an incipient charge density wave (CDW) is formed below $T^*$, which onsets at about $T_{CDW} \approx 130$ - 150 K in the doping range at $p = 0.11 - 0.14$ ( ref. 39,40). At a lower temperature $T_{FSR} \approx 50$ K, high-field NMR[12] and ultrasound[14] measurements for $p = 0.11$ have revealed a phase transition, below which long range static charge order appears and Fermi surface reconstruction is believed to take place. Taking into account that the $\rho(0)$ values shown in Fig. 3 are extrapolated from this high-field phase, there would appear to be no significant change in the $A$ coefficient between the low-$T$ regime where long range CDW sets in and the high-$T$ regime where only incipient CDW order is detected. This is reminiscent of the situation in NbSe$_2$ where there is negligible change of the resistivity at the CDW transition $T_{CDW} = 33$ K (ref. 41). This behavior can be understood if no substantial change of the Fermi surface occurs at the CDW transition (for a discussion see ref. 42). To make an analogy with Y124, we first acknowledge that the effect of the pseudogap is to suppress quasiparticles near the Brillouin zone boundaries. A Fermi surface reconstruction due to charge order with dominant wavevectors ($Q_x$, 0) and (0, $Q_y$) will create a small electron-like pocket composed of the residual 'nodal' density of states, in contrast to the large hole-like Fermi surface characterizing the overdoped state. This Fermi surface reconstruction can be seen as a folding of the Fermi surface and in terms of transport properties, the same nodal states will be involved in scattering processes below $T^*$ and below $T_{FSR}$ when the Fermi surface is reconstructed. Thus, the similar value of $A$ at high and low temperatures can be reconciled. Note that in canonical 1D CDW systems such as NbSe$_3$ (ref. 43) and organics metals[44], there is a marked change of slope of the resistivity below the charge ordering temperature due to nesting of part of the original Fermi surface.

The scenario discussed above for Y124 is also consistent with the observation of an anisotropic scattering rate $\Gamma$ in cuprates. In overdoped $Tl_2Ba_2CuO_{6+\delta}$ (Tl2201) (ref. 45), for example, it has been shown that the scattering rate is composed of two distinct terms, a $T^2$ term that is almost isotropic within the basal plane and a $T$-linear scattering rate that is strongly anisotropic, vanishing along the zone diagonals and exhibiting a maximum near the Brillouin zone boundary, where the pseudogap is maximal. Inside the pseudogap regime, therefore, one expects the $T$-linear scattering rate to become much diminished, leaving the $T^2$ scattering term as the dominant contribution to $\rho_a(T)$.

In correlated metals, both the $T^2$ resistivity and the $T$-linear specific heat are the consequence of the Pauli exclusion principle. Thus, the strength of the $T^2$ term in $\rho(T)$ is empirically related to the square of the electronic specific heat coefficient $\gamma_0$ via the Kadowaki-Woods ratio (KWR) $A/\gamma_0^2$ (ref. 46). An explicit expression for the KWR has been derived for correlated metals taking into account unit cell volume, dimensionality and carrier density[47]. In a single band quasi-2D metal, the $A$ coefficient reads:

$$A_{KWR} = \left(\frac{8\pi^3 ac k_B^2}{e^2\hbar^3}\right) \cdot \left(\frac{m^{*2}}{k_F^3}\right) \qquad (2)$$

where $a$ and $c$ are the lattice parameters, and $m^*$ and $k_F$ are respectively the (isotropic) effective mass and Fermi wavevector. A similar expression has also been obtained recently using the Kubo formalism[48]. As shown in the Supplementary, comparison of $A_{KWR}$ with experimentally determined values for both single- and multi-band correlated oxides shows good agreement for $A_{KWR}$ values spanning over three orders of magnitude.

Electron-electron collisions involve two quasiparticles that reside within a width of order $k_B T$ near the Fermi energy, providing the factor $T^2$. However, the total electron momentum is conserved in normal electron-electron scattering for the simple metals with a nearly-free-electron-like Fermi surface. Additional mechanisms, such as Umklapp or interband scattering, are thus needed in order to understand the dissipation (see ref. 49-51 for a discussion). With regards the hole-doped cuprates, it is debatable whether the KWR should hold at all within the pseudogap phase. However, given the increasing amount of data pointing to a rather conventional state at sufficiently low temperature, we believe that such analysis and comparison is appropriate and in the following, we consider briefly a number of these possible mechanisms[47-51] in turn.

### *Umklapp scattering*

Assuming a single electron pocket per $CuO_2$ plane, Umklapp collisions can lead to dissipation for electron-electron scattering since the condition on the reconstructed FS, $k_F > G/4$, is fulfilled in the reconstructed Brillouin zone ($G$ is a reciprocal lattice vector). In Y124, the QO frequency linked to the electron pocket ($F_e = 660 \pm 30$ T) converts into $k_{Fe} = 1.42 \pm 0.03$ nm$^{-1}$ (ref. 52,53), while the most recent QO measurements have indicated that $m^* = 1.9 \pm 0.1$ $m_e$ (ref. 54-55). Equation (2) thus gives an estimate of the $A$ coefficient for the electron pocket, $A_e = 86 \pm 20$ n$\Omega$ cm K$^{-2}$, i.e. $A_{KWR} = 43 \pm 10$ n$\Omega$ cm K$^{-2}$ taking into account the two $CuO_2$ planes. Significantly, this is almost five times larger than the experimental value. Assuming that the KWR ratio holds in underdoped cuprates, it implies that a reconstructed Fermi surface containing only one electron pocket (with an isotropic $m^*$) cannot account fully for the magnitude of the $T^2$ resistivity term in Y124. Similar conclusions are also drawn from comparison of the Fermi parameters reported for underdoped Y123 (ref. 18) and the measured $A$ coefficient[33] (albeit at elevated temperatures).

*Multiband scenario*

In a second scenario initially proposed by Baber, the momentum transfer between two distinct reservoirs can also lead to dissipative scattering[49]. In underdoped Y123, a small QO frequency ($F_h \approx 95$ T) was discovered and attributed to a hole pocket based on the doping dependence of the Seebeck coefficient[26]. A Fermi surface comprising at most one electron and two hole pockets with the measured areas and masses is consistent with a scenario based on the Fermi-surface reconstruction induced by the CDW order observed in Y123[24] (see Fig. 4). It is also compatible, within error bars, with the value of the electronic specific heat measured at high fields in YBCO[56,57] (see Supplementary for details). (Before proceeding, it is worth noting that the multiband scenario relies on the assumption that the small oscillations detected in ref. 26 derive from a small hole pocket. Other studies have suggested that the small oscillation $F_h$ owes its origin in quantum interference or the Stark effect[29] in a magnetic breakdown scenario between the bilayer of Y123 (ref. [23, 27, 28])).

Taking the parameters for the two types of pockets in underdoped Y123 ($p = 0.11$): $k_{Fe} = 1.28 \pm 0.02$ nm$^{-1}$ and $m^*_e = 1.4 \pm 0.1$ $m_e$, $k_{Fh} = 0.54 \pm 0.03$ nm$^{-1}$ and $m^*_h = 0.45 \pm 0.1$ $m_e$, we obtain $A_e = 56 \pm 9$ nΩ cm K$^{-2}$ and $A_h = 89 \pm 42$ nΩ cm K$^{-2}$ (note that the large error here is due to the relative error in the effective mass). Finally, applying the parallel-resistor formula and taking into account the bilayer nature of Y123, the $A$ coefficient is estimated to be $A_{KWR} \approx 12 \pm 6$ nΩ cm K$^{-2}$, in reasonable agreement with the value measured at high temperature ($A \approx 6.6$ nΩ cm K$^{-2}$) (ref. 33).

CDW order has not yet been directly observed by X-ray or by NMR in Y124, but the similar QO frequency (presumed to arise from the electron pocket) and the sign change in the Hall coefficient observed in both families[58] point to a very similar Fermi surface reconstruction. Assuming the presence of additional (though as yet undetected) hole pockets in the reconstructed Fermi surface of Y124 and given that $A = 9.5 \pm 1.5$ nΩ cm K$^{-2}$, we can estimate using Equation (2) the effective magnitude of the $A$ coefficient associated with an individual hole pocket to be $A_h = 49 \pm 10$ nΩ cm K$^{-2}$. Given the strong sensitivity of $A_{KWR}$ to the absolute values of $m^*$ and $k_F$, this estimate is considered to be in good agreement with the value of $A_h$ deduced for Y123. This is also in agreement with the two-band description of transport data in Y124[35] and the doping dependence of the Seebeck[20] and Hall[19] coefficients in YBCO. The magnitude of the $T^2$ term in $\rho_a(T)$ can thus be considered as yet further evidence that the reconstructed Fermi surface of underdoped Y123 and Y124 contains not only the well-established electron pocket, but also at least one additional hole-like pocket. This is in agreement with the Fermi surface reconstruction scenario proposed within the biaxial CDW model[24,25], assuming that the initial FS is the pseudogapped FS (e.g. without the states at the anti-node) and not the band structure derived FS.

The multi-band scenario has a number of other implications. Firstly, in the Hg1201 family, QOs have been measured at a doping level $p = 0.09$ with a frequency $F = 840 \pm 30$ T and an effective mass $m^* = 2.45 \pm 0.15$ $m_e$ (ref. 59). Accordingly, Equation (2) gives $A_{KWR} = 73 \pm 10$ nΩ cm K$^{-2}$. Above $T_c$, the measured $A$ coefficient is $A \approx 10$ nΩ.cm.K$^{-2}$ (ref. 34). This large discrepancy between the estimated and measured values of $A$ again indicates that the reconstructed Fermi surface of Hg1201 may also contain an additional pocket or pockets that have not yet been observed in QO experiments[60]. Secondly, in Y123, QOs have been observed over a wide range of doping yet $F$ is found to increase only by ~20 % between $p = 0.09$ and $p = 0.152$ (ref. 61). These measurements also reveal a strong enhancement of the quasiparticle effective mass as optimal doping is approached and suggest a quantum critical point at a hole doping of $p_{crit} \approx 0.17$. If the electron pocket was indeed the only pocket that persists in the reconstructed phase, this marked enhancement in $m^*$ should lead to a corresponding enhancement of the $A$ coefficient on both sides of $p_{crit}$, as has been observed, for example, in the transport properties of the isovalently substituted pnictide family BaFe$_2$(As$_{1-x}$P$_x$)$_2$ (ref. 38). In cuprates, however, the situation is far from clear. Certainly, there is no sign of a divergence of the $A$ coefficient near $p \approx 0.18$ from existing high temperature measurements (ref. 34). Moreover, measurements of the low-temperature in-plane resistivity of several overdoped LSCO samples in high magnetic field have revealed a regime of 'anomalous' or 'extended' criticality regime around $p \approx 0.19$ where the coefficient of the $T$-linear term is maximal yet there is no sign of divergence or an enhancement in $A$ (from the overdoped side)[5]. Within this multi-band scenario, it is possible that the marked enhancement in $m^*$ of the electron pocket, and thus in $A_e$, is offset by changes in $A_h$ (assuming that the hole pockets are always present).

### Alternative scenarios

Finally, we consider an alternative scenario for the pseudogap in which only Fermi arcs exist at low-field and ask what is the magnitude of the $A$ coefficient expected in such a scenario. This is relatively easy to do, at least approximately. For Y124, $p = 0.14$. Thus, the full Fermi surface is expected to occupy 57% of the Brillouin zone. If we assume that the quasiparticles on the full Fermi surface have a comparable mass to those found in overdoped cuprates, i.e. $m^* \sim 5$ $m_e$, one obtains $A_{KWR} = 2.5$ n$\Omega$ cm K$^{-2}$ for the full unreconstructed Fermi surface. At low-$T$, the 'normal state' specific heat coefficient $\gamma_0$ in Y124 is estimated using the entropy conservation construction to be approximately one third of its value at high temperature, i.e. above $T^*$ (ref. 52). For a Fermi arc that is one third the length of the full quadrant, $A_{KWR}$ is correspondingly tripled (again assuming an isotropic $m^*$). Thus, the magnitude of $A$ with such a scenario is similar to the coefficient found at low temperatures and from high field studies. It is important to recognize, however, that a Fermi arc has only hole-like curvature, and thus can in no way account for the two-carrier form of the magnetoresistance in Y124, nor for the negative sign of the Hall coefficient at low temperatures and high fields.

Until now, all estimates and comparisons have been made under the assumption that the effective mass does not vary around the Fermi surface. In an alternative scenario[62], the effective mass of the diamond-shaped electron pocket is taken to be anisotropic. The effective mass deduced from quantum oscillations is large because it is dominated by the corner of the pocket which corresponds to the hot spot of the CDW. By contrast, transport is dominated by those regions of the FS with the highest Fermi velocity, i.e. the light quasiparticles in the near-nodal state. This scenario can explain the factor of five discrepancy between the expected value of the KWR ratio and the experimental one (assuming a single electron pocket) but the anisotropy needs to be abnormally strong and would need to increase with doping in order to explain the behavior of the effective mass deduced from quantum oscillations[61]. Similar considerations would also apply if the scattering rate $\Gamma$, rather than the effective mass, were strongly anisotropic[63].

The picture of underdoped cuprates now emerging from X-ray spectroscopy, is of a lengthening of the correlation length $\xi$ associated with the charge ordering with increased field strength[15,16], in agreement with NMR and thermodynamic measurements. At what value of $\xi$, relative to the mean-free-path of the remnant quasiparticles, does Fermi surface reconstruction appear, or is manifest in the magneto-transport properties, is a crucial open question. At high temperatures, where the mean-free-path is short, the quasiparticle response will be susceptible even to short-range charge-order. At low temperatures, however, the situation is less clear. The findings reported in this manuscript point to a strong influence of the incipient charge order on the transport properties over a wide region of the temperature-doping-magnetic field phase diagram, and calls for a systematic study of the evolution of the low-$T$ in-plane resistivity of underdoped cuprates inside the pseudogap regime.

**Acknowledgments** We thank K. Behnia, M. Ferrero, A. Georges, and L. Taillefer for useful discussions. Research support was provided by the project SUPERFIELD of the French Agence Nationale de la Recherche, the Laboratory of Excellence "Nano, Extreme Measurements and Theory" in Toulouse, the High Field Magnet Laboratory–Radboud University/Fundamental Research on Matter, and Laboratoire National des Champs Magnétiques Intenses, members of the European Magnetic Field Laboratory.

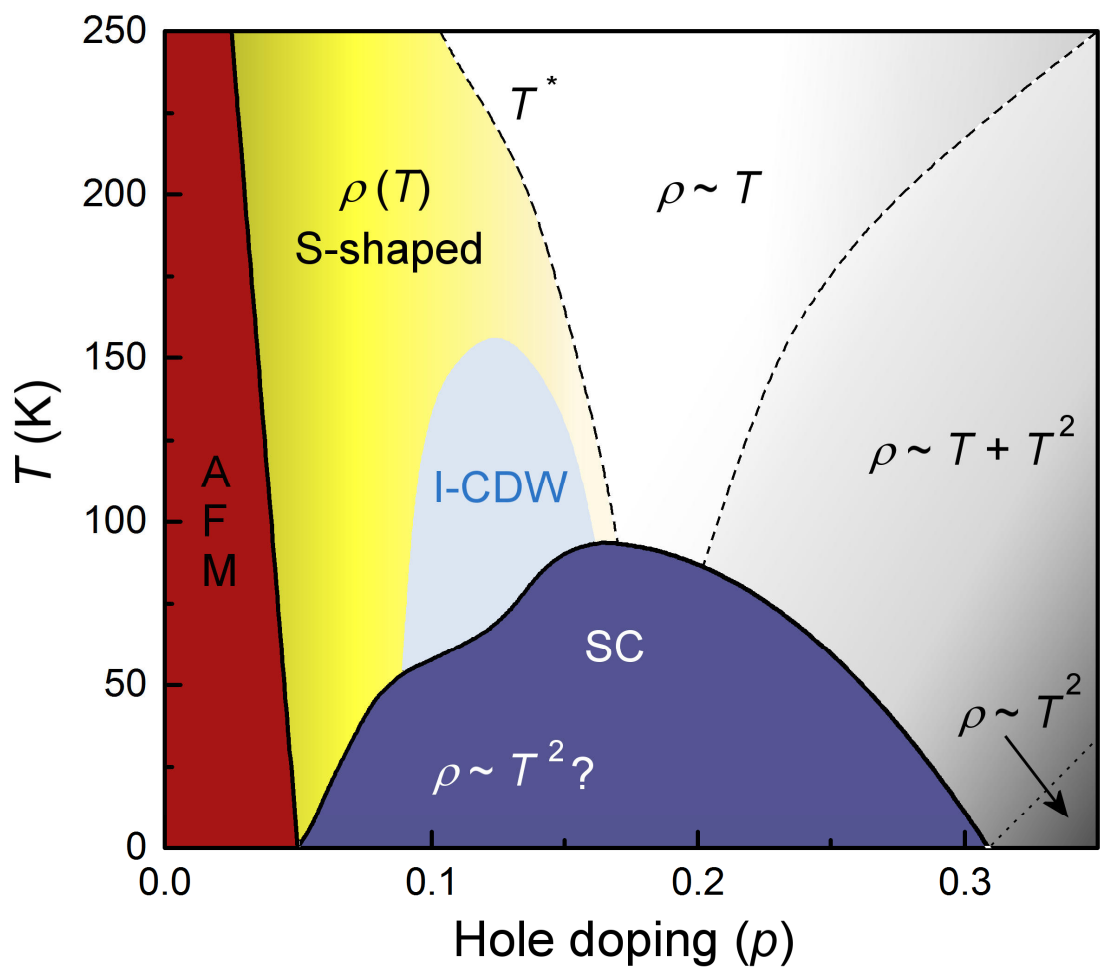

**Figure 1. Generic phase diagram of hole doped cuprates.** Generic phase diagram of hole-doped cuprates mapped out in terms of the temperature and doping evolution of the in-plane resistivity $\rho_{ab}(T)$. The solid lines are the phase boundaries between the normal state and the superconducting (SC) or antiferromagnetic (AFM) ground state. The dashed lines indicate crossovers in $\rho_{ab}(T)$ behaviour. The light blue area corresponds to where incipient charge order (I-CDW) has been detected by X-ray measurements in Y123[39,40]. The CDW becomes long range order below $T_c$ when a large magnetic field is applied[12-14].

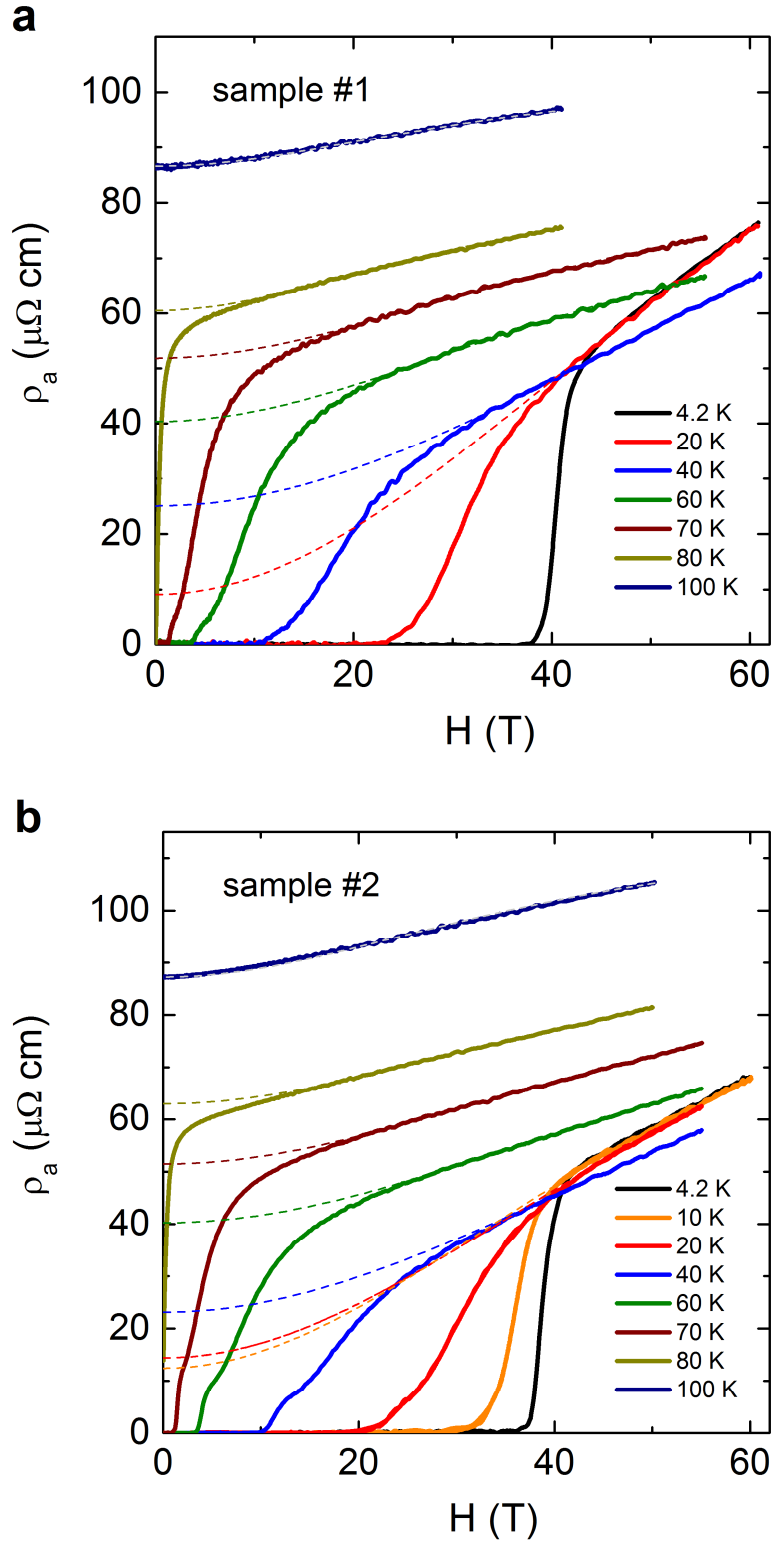

**Figure 2. Field dependence of the in-plane resistivity in Y124.** a,b) Electrical resistivity $\rho_a$ of two samples of YBa$_2$Cu$_4$O$_8$ ($T_c$ = 80 K) for current $I \parallel a$ and magnetic field $H$ applied along the $c$-axis at different temperatures (solid lines). With lowering the temperature, a strong magnetoresistance develops, which can be accounted for by a two-band model[35]. Dashed lines correspond to fits using this model (Equation (1) in the text) to extrapolate the normal-state data to $H$ = 0.

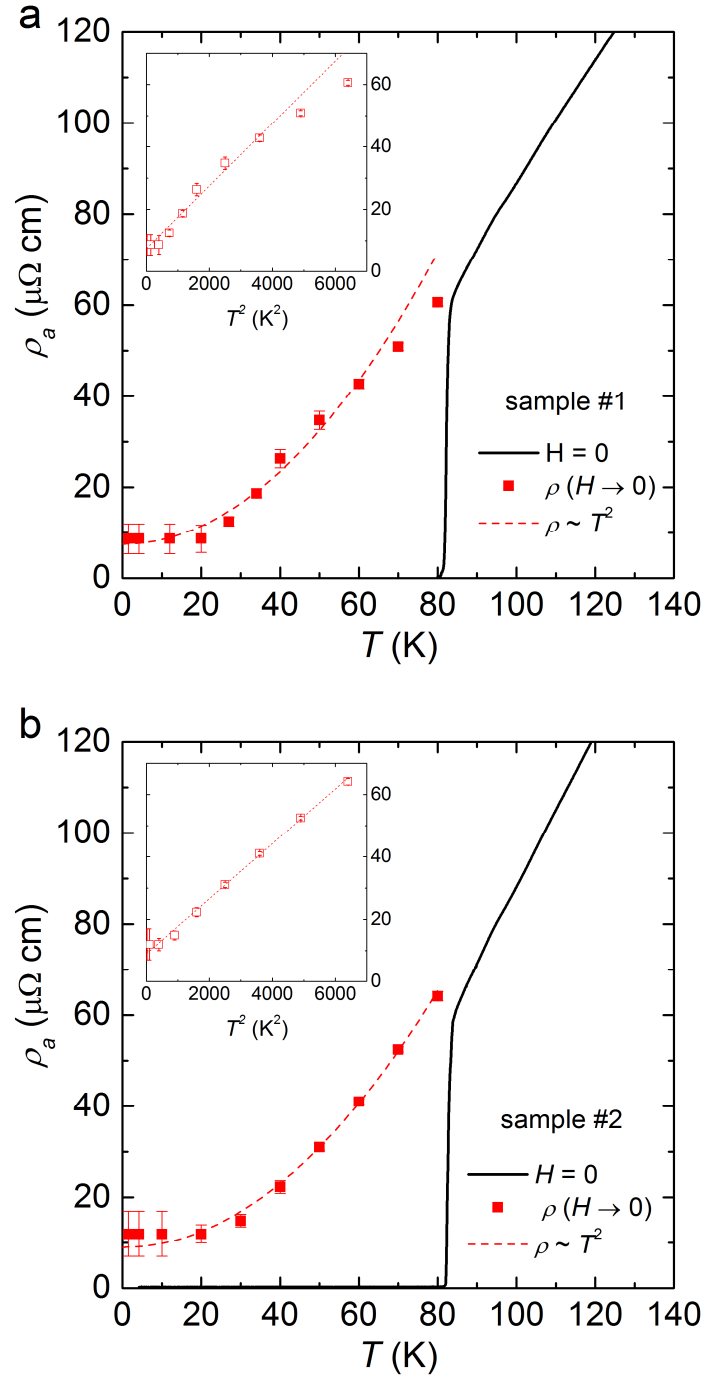

**Figure 3. Temperature dependence of the in-plane resistivity in Y124.** Temperature dependence of the $a$-axis resistivity of $YBa_2Cu_4O_8$ from which the magnetoresistance has been subtracted using a two-band model to extrapolate the normal-state data to $H = 0$ for the two samples shown in Fig. 2. Solid lines show the resistivity measured in zero magnetic field. Dashed lines are fits to the Landau-Fermi liquid expectation for the temperature dependence of the resistivity at low temperature, $\rho_a(T) = \rho_0 + AT^2$ (see text). To estimate error bars we fitted each field sweep data set to Equation (1) between a lower bound $H_{cutoff}$ and the maximum field strength and monitored the value of $\rho(0)$ as a function of $H_{cutoff}$ (see Supplementary for more details). The insets are corresponding plot of $\rho_a(T)$ versus $T^2$ to highlight the quadratic form of $\rho_a(T)$.

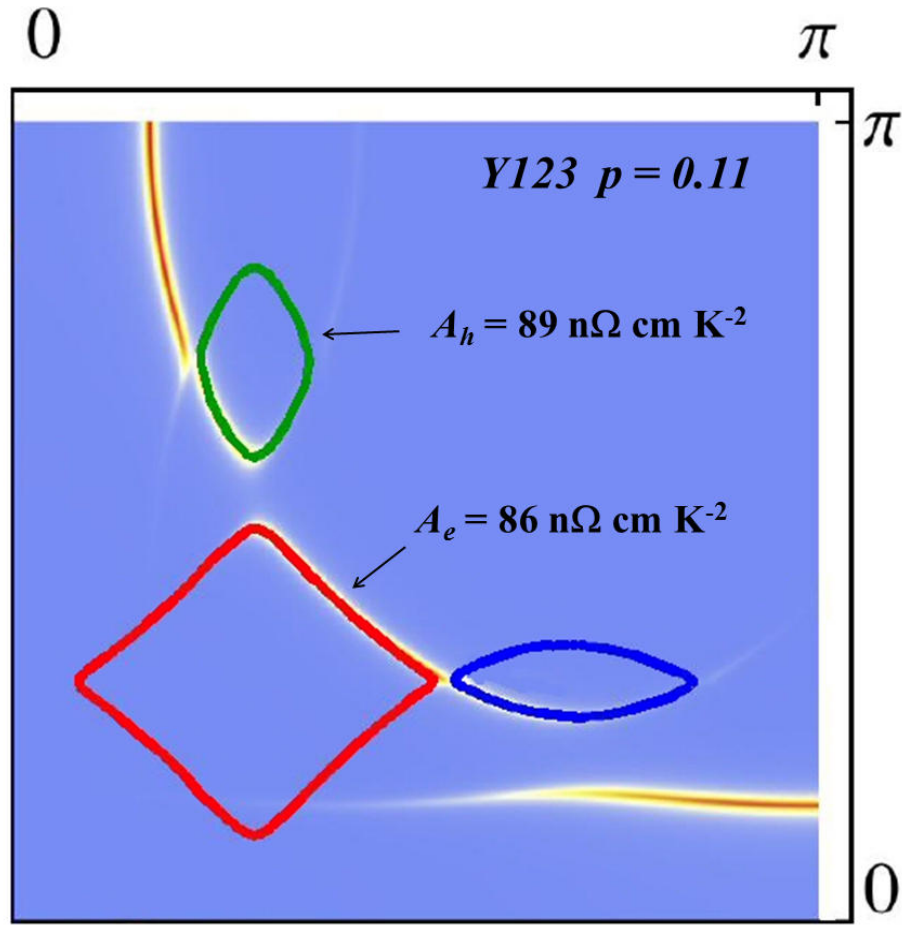

$$A_{tot} = \left[ 2 \left( \frac{1}{A_e} + \frac{2}{A_h} \right) \right]^{-1} = 14.7 \ n\Omega \ cm \ K^{-2}$$

**Figure 4. Fermi surface reconstruction by biaxial charge order.** Sketch of the reconstructed Fermi surface adapted from ref. 24 using the CDW wavevectors measured in Y123, showing a diamond-shaped nodal electron pocket (red) and two hole-like ellipses (blue and green). $A_e$ and $A_h$ are the $A$ coefficients estimated from Equation (2) for the electron and hole pockets, respectively. $A_{tot}$ is the expected $A$ coefficient for a parallel resistor model taking into account one electron and two hole pockets and the two $CuO_2$ planes.

# SUPPORTING INFORMATION

### 1. SAMPLES

Single crystalline samples (typical dimensions 400 x 80 x 30 $\mu m^3$) were flux grown in $Y_2O_3$ crucibles in a partial oxygen pressure of 400 bar [64]. In contrast to the Y123 family, which has a single CuO chain with variable oxygen content, Y124 contains alternating stacks of $CuO_2$ bilayers and double CuO chains (along $b$-axis) that are stable and fully loaded.

### 2. MEASUREMENT OF THE LONGITUDINAL AND TRANSVERSE RESISTANCES

The pulsed-field measurements were performed using a conventional 6-point configuration with a current excitation between 5 mA and 10 mA at a frequency of ~ 50 kHz. A high-speed acquisition system was used to digitize the reference signal (current) and the voltage drop across the sample at a frequency of 500 kHz. The data were post-analyzed using software to perform the phase comparison. Data for the rise and fall of the field pulse were in good agreement, thus excluding any heating due to eddy currents.

### 3. TWO-BAND MODEL

Assuming that the Fermi surface of underdoped $YBa_2Cu_3O_y$ contains both electron and hole pockets, the transverse magnetoresistance can be fitted with a two-band model:

$$\rho(H) = \frac{(\sigma_h + \sigma_e) + \sigma_h \sigma_e (\sigma_h R_h^2 + \sigma_e R_e^2) H^2}{(\sigma_h + \sigma_e)^2 + \sigma_h^2 \sigma_e^2 (R_h + R_e)^2 H^2} = \rho_0 + \frac{\alpha H^2}{1 + \beta H^2}$$

(S1)

Where $\rho_0 = \dfrac{1}{\sigma_h + \sigma_e}$ (S2)

$$\alpha = \frac{(\sigma_h + \sigma_e)\sigma_h \sigma_e (\sigma_h R_h^2 + \sigma_e R_e^2) - \sigma_h^2 \sigma_e^2 (R_h + R_e)^2}{(\sigma_h + \sigma_e)^3}$$

(S3)

$$\beta = \frac{\sigma_h^2 \sigma_e^2 (R_h + R_e)^2}{(\sigma_h + \sigma_e)^2}$$

(S4)

$\sigma_h$ ($\sigma_e$) is the conductivity of holes (electrons) and $R_h$ ($R_e$) is the Hall coefficient for hole (electron) carriers.

Using the three free fitting parameters $\rho_0$, $\alpha$ and $\beta$, we were able to subtract the orbital magnetoresistance from the field sweeps and get the temperature dependence of the extrapolated zero-field resistivity $\rho(0)$, which is the main focus of the present study. A detailed description and discussion of $\alpha$ and $\beta$, meanwhile, is considered beyond the scope of the present paper.

The choice of this model is justified by examining closely the form of the magnetoresistance at high temperatures above $T_c$. In Fig. S1, we compare the fit using the two-band model with a simple $H^2$ magnetoresistance. At $T$ = 100 K where there is no trace of superconductivity, it is evident that a simple $H^2$ magnetoresistance cannot fit the data whereas the two-band model provides a reasonable fit to the data over the entire field range.

## 4. FIT OF THE MAGNETORESISTANCE

To track the temperature dependence of the zero-field resistivity $\rho(0)$, we fitted each field sweep data set between some variable lower bound $H_{\text{cut-off}}$ and $H_{\text{max}}$, the peak field of each pulse, to the two-band form $\rho(H) = \rho(0) + \frac{\alpha H^2}{1 + \beta H^2}$ and monitored the value of $\rho(0)$ as a function of $H_{\text{cut-off}}$. Example plots of $\rho(0)$ versus each given $H_{\text{cut-off}}$ are shown in Figure S2 (open symbols). For $H_{\text{cut-off}}$ less than some critical field, defined as $H^*$, the extracted $\rho(0)$ values rise monotonically with increasing cut-off field as one enters the resistive transition. For $H_{\text{cut-off}} >$ $H^*$, $\rho(0)$ reaches a plateau value, indicating that the magnetoresistance now follows the two-band form shown above. As the field range for fitting becomes too narrow, the extrapolated value of $\rho(0)$ begins to oscillate wildly (not shown). Having determined $H^*$, we then fitted each field sweep data set between $H^*$ (now fixed) and some variable upper bound $H_{\text{upper}}$ (full symbols in Fig. S2 and S3).

At all temperatures, the field range to obtain reliable values of $\rho(0)$ is of the order of $10 - 15$ T or more concretely, between $1.3\,H^* \leq H \leq H_{\text{max}}$. Indeed, if we look to the curve at $T$ = 60 K ($H_{\text{max}}$ = 55 T), for example, the value of $\rho(0)$ changes only by a few percent when changing the upper cut-off from 55 T down to 40 T. In another words, fitting the data between $H^*$ = 30 T and $H_{\text{upper}}$ = 40 T gives the same value of $\rho(0)$ to within $\pm$ 5 % as fitting between 30 T and 55 T. We used the scatter in the value of $\rho(0)$ when fitting the data with a upper cut-off to define the error bar (standard deviation).

At temperatures lower than $T$ = 20 K, we noticed that all curves merge on top on the others for sample #1 (see Fig. S4) showing that the magnetoresistance and the extrapolated resistivity at $H$ = 0 cannot be differentiated within our experimental uncertainty. For sample #2, due to a smaller maximum field at T = 20 K and to a slight temperature dependence of the magnetoresistance, the error bars are higher below T = 20 K.

## 5. EFFECTIVE MASS AND SPECIFIC HEAT

From the measured effective mass $m^*$, the residual linear term $\gamma$ in the electronic specific heat $C_e(T)$ at $T \rightarrow 0$ can be estimated through the relation [65]

$$\gamma = (1.46 \text{ mJ / K}^2 \text{ mol}) \, \Sigma_i \, (n_i \, m_i^* / m_0)$$

where $n_i$ is the multiplicity of the $i^{\text{th}}$ type of pocket in the first Brillouin zone (This expression assumes an isotropic Fermi liquid in two dimensions with a parabolic dispersion). For a Fermi surface containing one electron pocket and two hole pockets per $CuO_2$ plane, we obtain a total mass of $(1.4 \pm 0.1) + 2\,(0.45 \pm 0.1) = 2.3 \pm 0.3\,m_0$, giving $\gamma = 6.7 \pm 0.9$ mJ / K$^2$ mol (for two $CuO_2$ planes per unit cell). High-field measurements of $C_e$ at $T \rightarrow 0$ in YBCO at $p \sim 0.11$ yield $\gamma = 6.5$

± 1.5 mJ / K$^2$ mol [57] at $H > H_{c2}$ = 30 T. Note that this measured value includes a residual electronic specific heat contribution that is present even in zero-field. We therefore find that the Fermi surface of YBCO can contain at most two small hole pockets in addition to only one electron pocket per CuO$_2$ plane. No further sheet can realistically be present in the Fermi surface.

## 6. KADOWAKI-WOODS ESTIMATION IN CORRELATED SYSTEMS

Comparison of $A_{KWR}$ with experimentally determined values shows good agreement for both single- and multi-band correlated systems. In the single-band, highly overdoped LSCO ($p$ = 0.33) for example, $A$ = 2.5 ± 0.5 nΩ cm.K$^{-2}$ [2]. This compares with $A_{KWR}$ = 4 ± 1 nΩ cm.K$^{-2}$ (using $m^*$ = 4.7 ± 1.0 $m_e$ and $k_F$ = 5.5 ± 0.2 nm$^{-1}$ from specific heat [2] and ARPES measurements [66]). Similarly, in highly overdoped Tl2201 ($p$ = 0.3), $A$ = 5.4 ± 0.5 nΩ cm.K$^{-2}$ [67], while from detailed quantum oscillation measurements ($m^*$ = 5.2 ± 0.5 $m_e$ and $k_F$ = 7.4 ± 0.1 nm$^{-1}$ [68]), one obtains $A_{KWR}$ = 3.9 ± 0.4 nΩcm.K$^{-2}$.

In multi-band systems, one may assume that bands contribute in parallel, e.g. $1/A = \sum_i 1/A_i$. In Sr$_2$RuO$_4$, for example, $A$ = 5 ± 1.5 nΩ cm.K$^{-2}$ [69]. Taking the values for $m^*$ and $k_F$ for the three Fermi cylinders deduced from QO measurements [70], one obtains $A_{KWR}$ = 3.6 ± 0.5 nΩ cm.K$^{-2}$, again in good agreement with the measured value. In isovalently substituted Ca$_{2-x}$Sr$_x$RuO$_4$, the system goes from a three-band metal (at $x$ = 2) to a single-band metal in the Ca-free case for $x$ < 0.5 as the so-called $\alpha$ and $\beta$ bands become localized [71]. This crossover from a single- to a multi-band system leads to an increase in the KWR of three orders of magnitude [71] that can be explained quantitatively using Equation (2) of the main text.

With regards to the hole-doped cuprates, it is questionable whether the KWR should hold at all within the pseudogap phase. However, given the increasing amount of data pointing to a rather conventional state at sufficiently low temperature, we believe that such analysis and comparison is appropriate. Thus, we are confident that the approach described here is also applicable to the reconstructed phase of underdoped cuprates.

**Table S1:** List of previous reports of T$^2$ resistivity in underdoped cuprates

| Compound | Doping level | T-range of T$^2$ resistivity | Reference |
|---|---|---|---|
| YBCO | 0.03 | 140 K to 300 K | [36] |
| YBCO | 0.09 | 60 K to 150 K | [33] |
| Hg1201 | 0.055 | 100 K to 190 K | [34] |
| Hg1201 | 0.075 | 85 K to 220K | [34] |
| Hg1201 | 0.1 | 90 K to 170K | [34] |
| LSCO | 0.01 | 180 K to 300 K | [34] |
| LSCO | 0.02 | 140 K to 250 K | [36] |
| LSCO | 0.08 | 60 K to 160 K | [36] |

**Table S2:** Values and error bars of the parameters of the two-band model for sample #1:

| Temperature[K] | $\rho(0)$ [$\mu\Omega$ cm] | $\alpha$ [$10^{-3}$ $\mu\Omega$ cm T$^{-2}$] | $\beta$ [$10^{-3}$ T$^{-2}$] |
|---|---|---|---|
| 80 | $60.6 \pm 0.8$ | $23.8 \pm 7$ | $1.1 \pm 0.5$ |
| 70 | $50.9 \pm 0.8$ | $20.75 \pm 3$ | $0.64 \pm 0.12$ |
| 60 | $42.7 \pm 1.0$ | $14.4 \pm 2.3$ | $0.3 \pm 0.08$ |
| 50 | $34.7 \pm 1.4$ | $14.3 \pm 2.3$ | $0.2 \pm 0.07$ |
| 40 | $26.3 \pm 2.0$ | $16.7 \pm 3$ | $0.14 \pm 0.06$ |
| 34 | $18.7 \pm 1.0$ | $22.7 \pm 1.3$ | $0.17 \pm 0.02$ |
| 27 | $12.5 \pm 1.0$ | $27.4 \pm 1.1$ | $0.17 \pm 0.01$ |
| 20 | $8.7 \pm 3.0$ | $33 \pm 3$ | $0.22 \pm 0.03$ |
| 12 | $8.7 \pm 3.0$ | $33 \pm 3$ | $0.22 \pm 0.03$ |
| 4.2 | $8.7 \pm 3.0$ | $33 \pm 3$ | $0.22 \pm 0.03$ |
| 1.5 | $8.7 \pm 3.0$ | $33 \pm 3$ | $0.22 \pm 0.03$ |

**Table S3:** Values and error bars of the parameters of the two-band model for sample #2:

| Temperature[K] | $\rho(0)$ [$\mu\Omega$ cm] | $\alpha$ [$10^{-3}$ $\mu\Omega$ cm T$^{-2}$] | $\beta$ [$10^{-3}$ T$^{-2}$] |
|---|---|---|---|
|  | $64.1 \pm 1.0$ | $12.8 \pm 1.4$ | $0.36 \pm 0.07$ |
| 70 | $52.5 \pm 0.6$ | $13.3 \pm 0.7$ | $0.28 \pm 0.03$ |
| 60 | $41.0 \pm 0.6$ | $13.7 \pm 0.6$ | $0.22 \pm 0.02$ |
| 50 | $31.0 \pm 0.8$ | $14.8 \pm 0.7$ | $0.19 \pm 0.02$ |
| 40 | $22.3 \pm 1.4$ | $19.7 \pm 1.1$ | $0.22 \pm 0.02$ |
| 30 | $14.9 \pm 1.4$ | $26.3 \pm 1.1$ | $0.26 \pm 0.01$ |
| 20 | $12 \pm 2$ | $30 \pm 5$ | $0.29 \pm 0.05$ |
| 10 | $12 \pm 5$ | $30 \pm 10$ | $0.29 \pm 0.1$ |
| 4.2 | $12 \pm 5$ | $30 \pm 10$ | $0.29 \pm 0.1$ |
| 1.5 | $12 \pm 5$ | $30 \pm 10$ | $0.29 \pm 0.1$ |

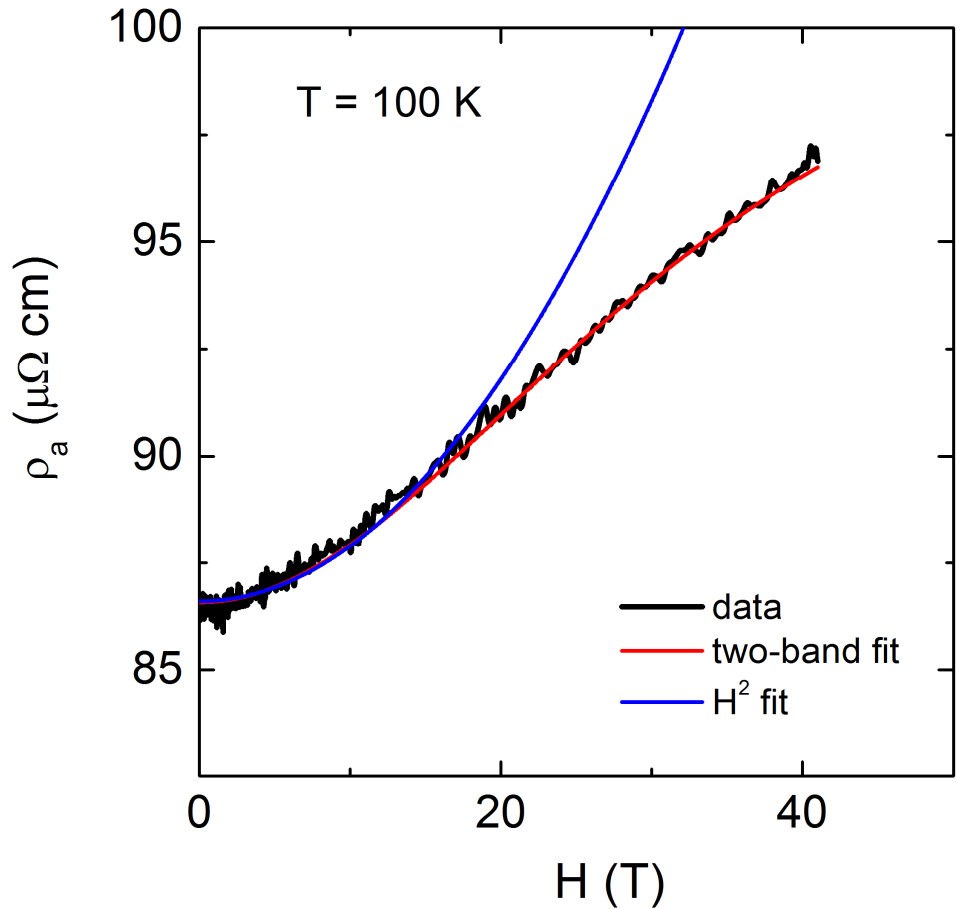

**Figure S1: Comparison of the fit using a two-band model and a $H^2$ magnetoresistance at $T$ = 100 K (sample #1).**

Electrical resistivity $\rho_a$ of YBa$_2$Cu$_4$O$_8$ (sample #1) at $T$ = 100 K (black line). The red curve corresponds to a fit to the data using the two-band expression $\rho(H) = \rho(0) + \frac{\alpha H^2}{1 + \beta H^2}$. For the blue curve, the low-field data were fitted to the single-band form $\rho(H) = \rho(0) + aH^2$.

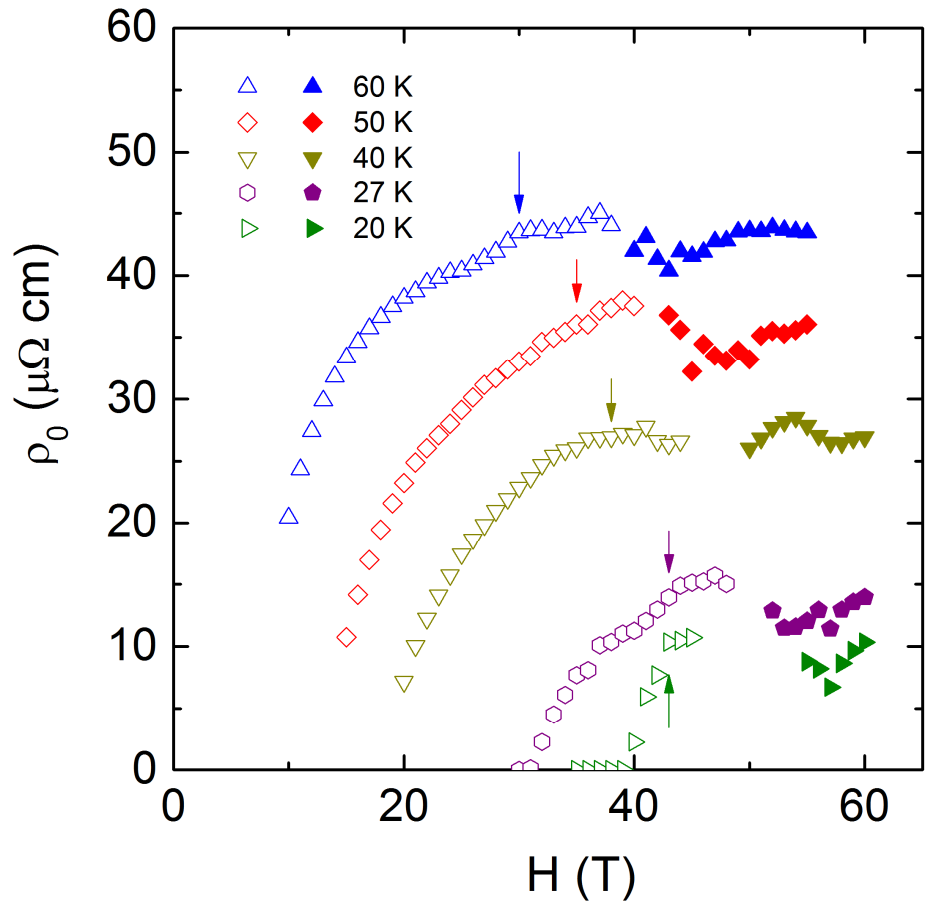

**Figure S2: Mode of extraction of the zero field resistivity ρ(H = 0) for Y124 (sample #1) at different temperatures.**

The data were first fitted to the form $\rho(H) = \rho(0) + \frac{\alpha H^2}{1 + \beta H^2}$ between some lower bound $H_{\text{cut-off}}$ and the peak field of each pulse and a value for $\rho(0)$ extracted (open symbols). The arrows indicate the value of the field strength $H^*$ below which the extrapolated $\rho(0)$ value decreases rapidly as one enters the resistive transition. The data were then fitted to the same expression between $H^*$ and some upper bound $H_{\text{upper}}$ which is itself then varied. The values of $\rho(0)$ deduced from this procedure (full symbols) are shown plotted as a function of $H_{\text{upper}}$, the sample held at the various temperatures indicated. These values indicated by the solid symbols are then averaged to determine $\rho(0)$ for each pulse.

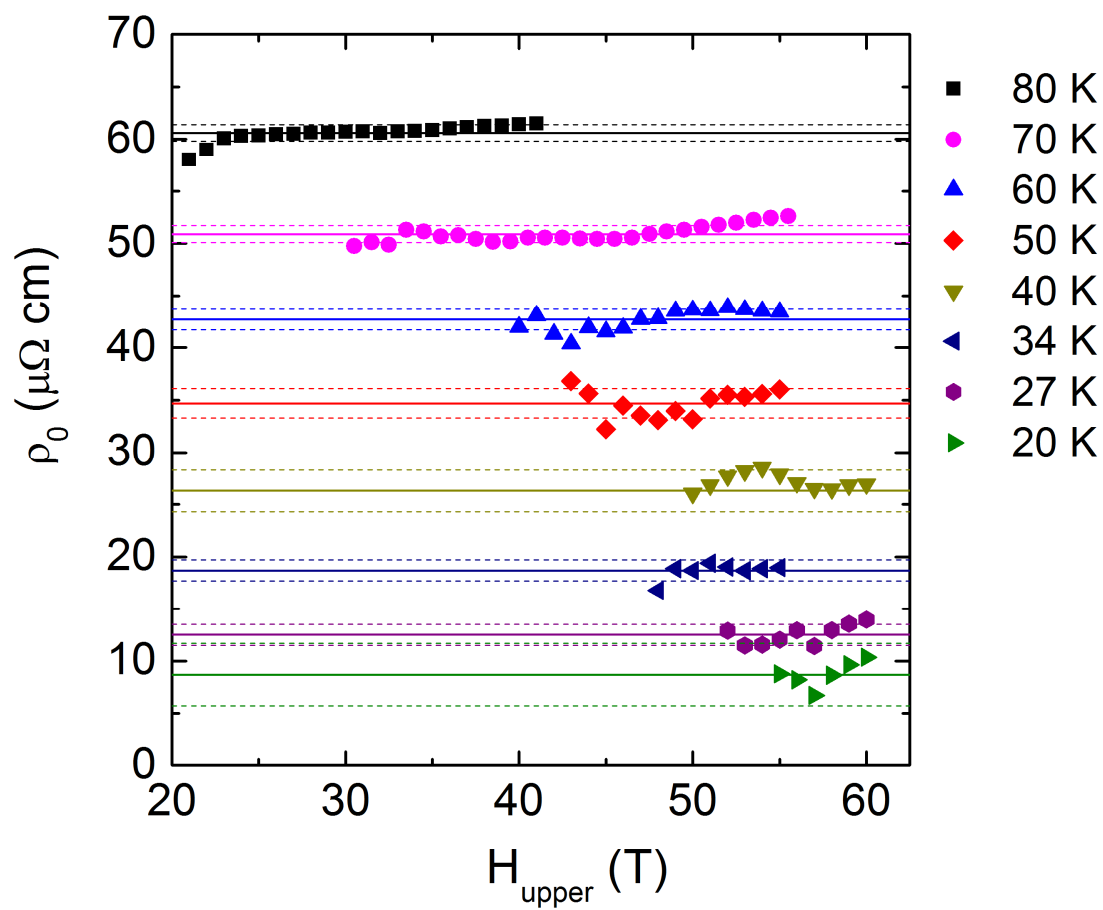

**Figure S3: Estimation of the error bars of *ρ(0) for* sample #1**

Same as Fig. S2 with only the extrapolated $\rho(0)$ values obtained from the fit between $H^*$ and $H_{upper}$ shown for clarity. The solid horizontal lines mark the mean value for $\rho(0)$ at each temperature while the dashed lines indicate the corresponding error bars.

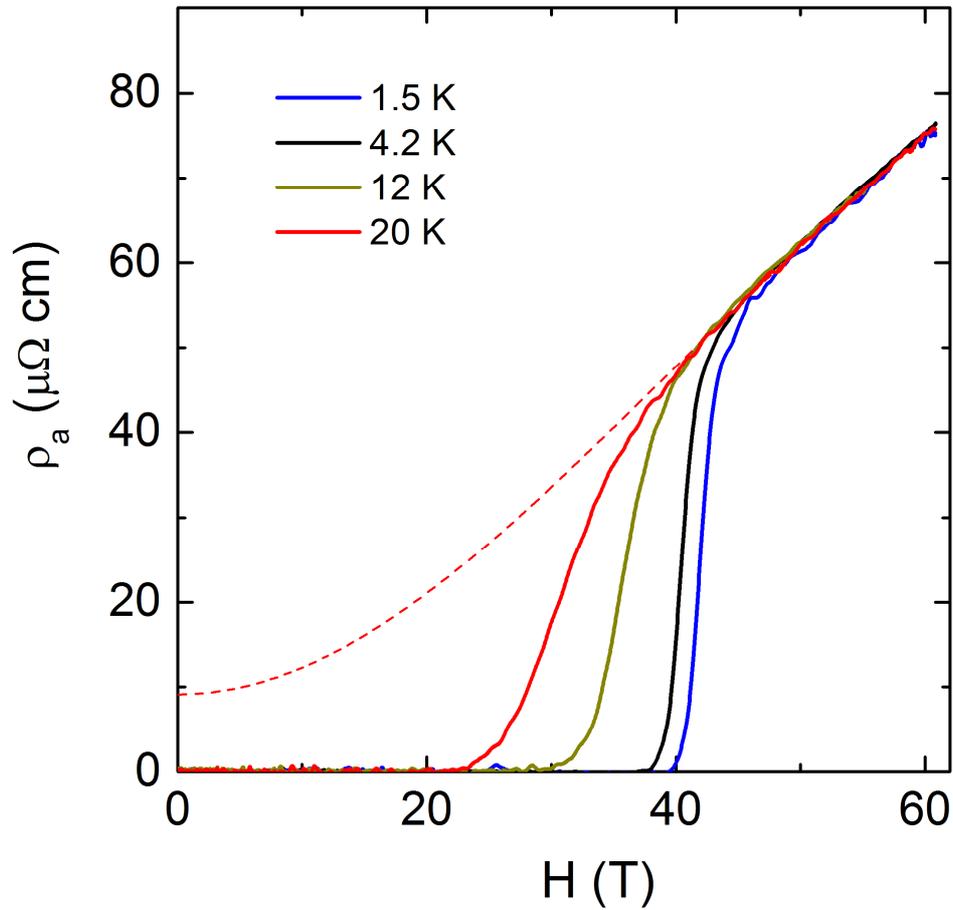

**Figure S4: Field dependence of the in-plane resistivity in Y124 (sample #1) at low temperatures.**

Electrical resistivity $\rho_a$ of $YBa_2Cu_4O_8$ (sample #1) between $T = 1.5$ K and $T = 20$ K, showing that the magnetoresistance and the extrapolated resistivity at $H = 0$ cannot be differentiated below 20 K within our experimental uncertainty. The dashed line corresponds to the two-band fit at $T = 20$ K used to extrapolate the normal-state data to $H = 0$.

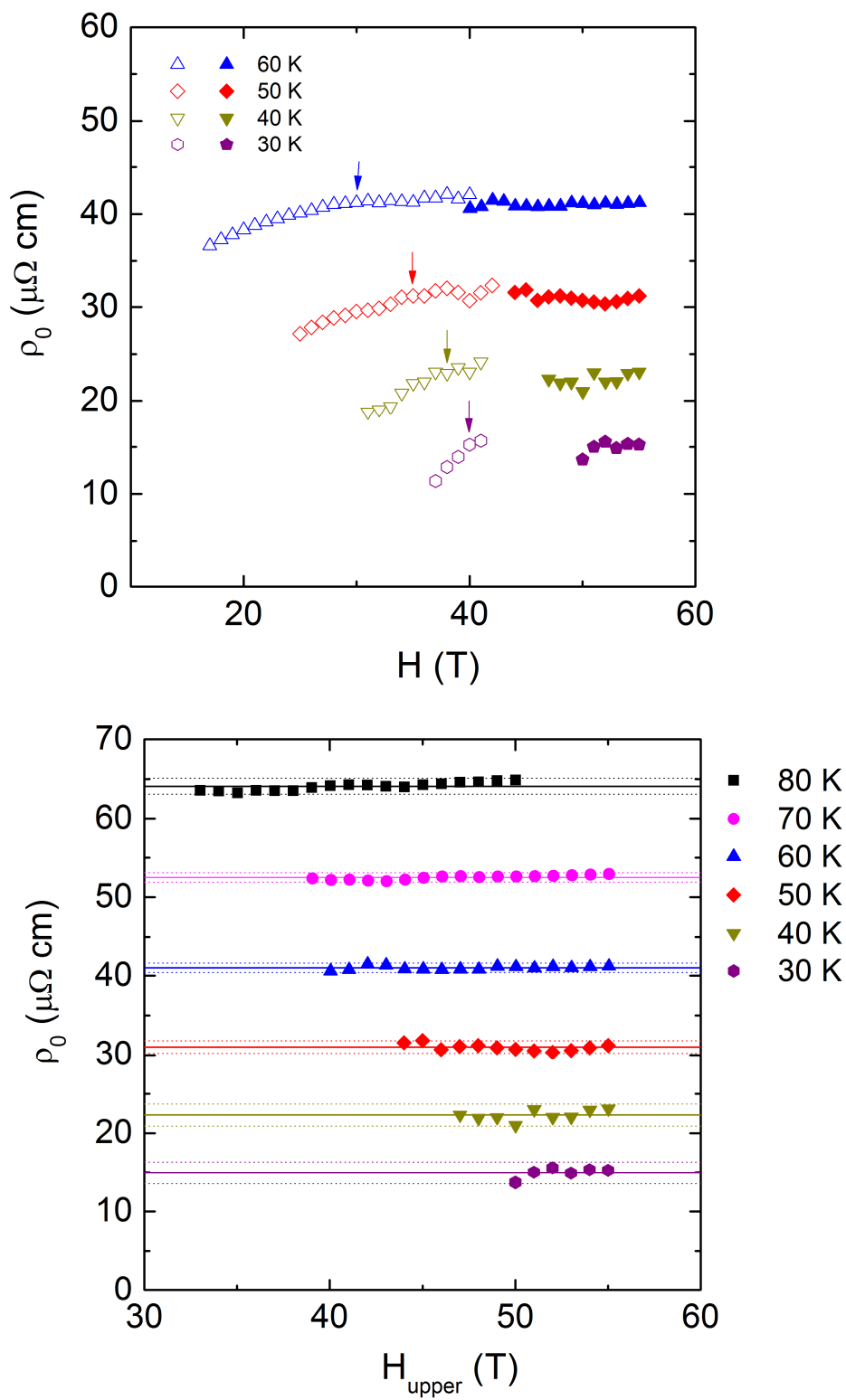

**Figure S5: Mode of extraction of the zero field resistivity $\rho(0)$ and error bars for sample #2 at the different temperatures indicated.**

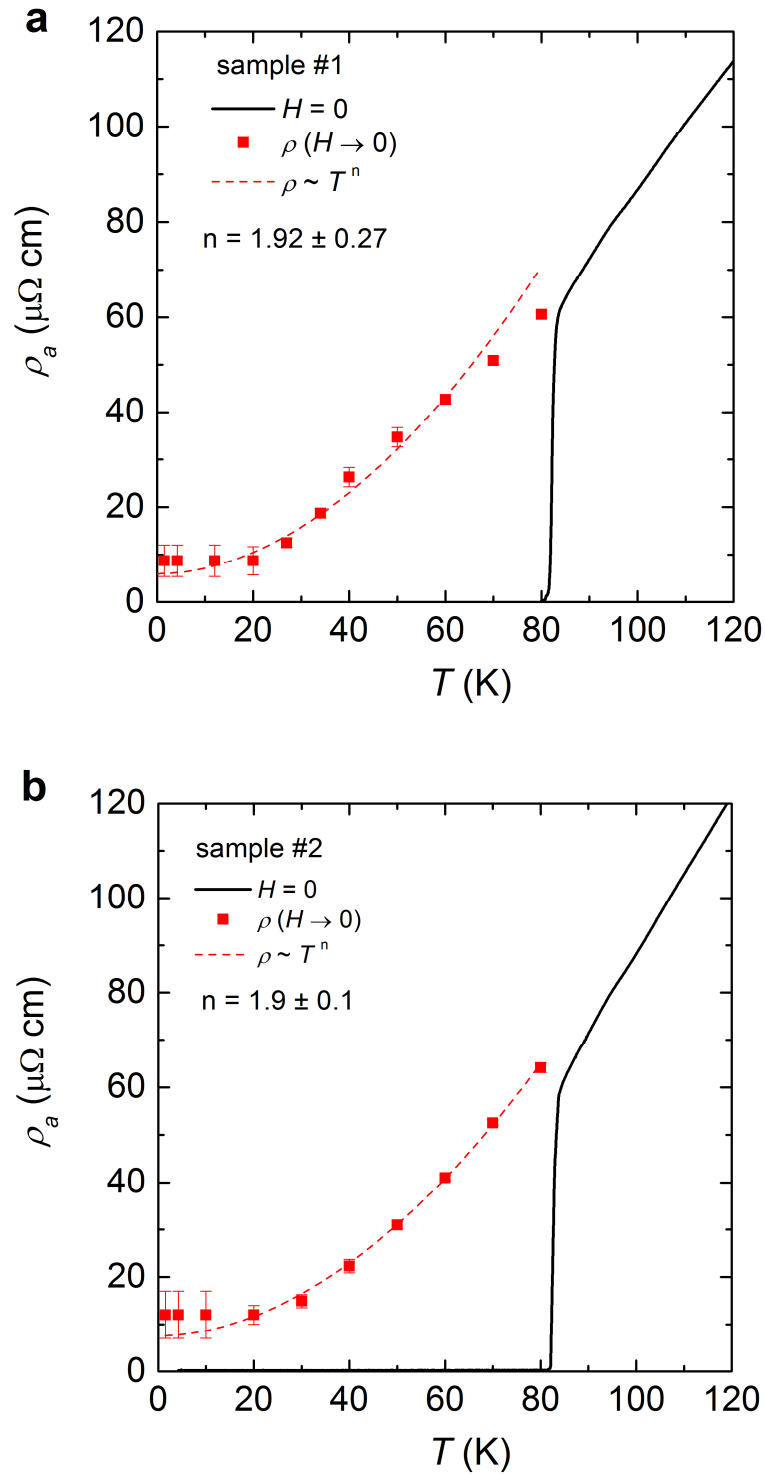

**Figure S6: Fits of the resistivity to a power law $\rho_a(T) = \rho_0 + AT^n$.**

Temperature dependence of the $a$-axis resistivity of $YBa_2Cu_4O_8$ from which the magnetoresistance has been subtracted using a two-band model to extrapolate the normal-state data to $H = 0$ for the two samples shown in Fig. 2. Solid lines show the resistivity measured in zero magnetic field. Dashed lines are fits to a power law $\rho_a(T) = \rho_0 + AT^n$ where the exponent $n$ is indicated in the figure for each sample. Note that the points above 60 K have been omitted for the fit of sample #1.